\documentclass[10pt,aps,prl,twocolumn,superscriptaddress,preprintnumbers]{revtex4-1}
\pdfoutput=1
\usepackage{amsfonts}
\usepackage{mathrsfs}
\usepackage{amsmath}
\usepackage{amssymb}
\usepackage{fancyhdr}
\usepackage{graphicx}
\usepackage{xspace}
\usepackage{rotating}
\usepackage[normalem]{ulem}
\usepackage{braket}
\usepackage{verbatim}
\usepackage{xcolor}
\usepackage[utf8]{inputenc}
\usepackage[medium]{titlesec}
\usepackage{bm}
\usepackage[normalem]{ulem}
\usepackage{extarrows}
\usepackage{slashed}
\usepackage{isodateo}
\usepackage{graphicx}
\usepackage{xcolor}
\usepackage[bookmarksnumbered=true,bookmarksopen=true]{hyperref}
\usepackage[hmargin=.7in,vmargin=1.1in]{geometry}
\usepackage{indentfirst}
\usepackage{cancel}
\usepackage{soul}

\graphicspath{{./}{./Figs/}{./figs/}{../Figs/}{./fig/}}

\usepackage{amsfonts}
\usepackage{mathtools}
\usepackage{pifont}
\usepackage{mathrsfs}
\usepackage{amsmath}
\usepackage{amssymb}
\usepackage{framed}

\newcommand{\bit}{\begin{itemize}}  
\newcommand{\eit}{\end{itemize}}

\usepackage{accents}

\newcommand{\benum}{\begin{enumerate}}
\newcommand{\eenum}{\end{enumerate}}
\newcommand{\bi}{\begin{itemize}}
\newcommand{\ei}{\end{itemize}}

\newcommand{\beq}{\begin{equation}}
\newcommand{\eeq}{\end{equation}}

\newcommand{\bea}{\begin{eqnarray}}
\newcommand{\eea}{\end{eqnarray}}

\newcommand{\Rmnum}[1]{\expandafter\@slowromancap\romannumeral #1@}

\def\bga{\begin{aligned}}
\def\eda{\end{aligned}}
\def\bgp{\begin{pmatrix}}
\def\edp{\end{pmatrix}}
\def\bgs{\begin{subequations}}
\def\eds{\end{subequations}}

\renewcommand{\rm}{\mathrm}

\usepackage{eso-pic}

\begin{document}

\preprint{NORDITA-2026-101}

\title{Primordial turbulence from inflation: a new inflaton-driven turbulent regime}
%\\

\author{Oksana Iarygina}
\email{oksana.iarygina@su.se}
\affiliation{Nordita, KTH Royal Institute of Technology and Stockholm University, Hannes Alfv\'ens v\"ag 12, 10691 Stockholm, Sweden}
\affiliation{The Oskar Klein Centre, Stockholm University, 10691 Stockholm, Sweden}
\affiliation{Instituut-Lorentz of Theoretical Physics, Universiteit Leiden, 2333 CA Leiden, The Netherlands}

\author{Axel Brandenburg}
\email{brandenb@nordita.org}
\affiliation{Nordita, KTH Royal Institute of Technology and Stockholm University, Hannes Alfv\'ens v\"ag 12, 10691 Stockholm, Sweden}
\affiliation{The Oskar Klein Centre, Stockholm University, 10691 Stockholm, Sweden}
\affiliation{McWilliams Center for Cosmology \& Department of Physics, Carnegie Mellon University, Pittsburgh, PA 15213, USA}
\affiliation{School of Natural Sciences and Medicine, Ilia State University, 3-5 Cholokashvili Avenue, 0194 Tbilisi, Georgia}

%
%\date{\today}

\begin{abstract}
%For the first time, we establish a direct connection between inflationary dynamics and the subsequent magnetohydrodynamic (MHD) evolution of primordial fields.
For the first time, we establish a direct connection between inflationary dynamics and the subsequent magnetohydrodynamic 
evolution. Starting from vacuum fluctuations,
we self-consistently evolve the coupled inflaton--plasma system through inflation, reheating, and into the radiation-dominated era. We find an inflaton-driven inverse cascade followed by freely decaying turbulence, with kinetic energy dominating over magnetic energy and driving dynamo amplification.
The inflaton-driven turbulence causes steeper magnetic energy decay than helical turbulence and should therefore be accounted for subsequent evolution of primordial fields and their present-day observables.

\end{abstract}
\maketitle

\noindent\textbf{Introduction.} \\
The early Universe provides a unique laboratory for studying how quantum fluctuations evolve into macroscopic fields. During inflation, quantum fluctuations are stretched to cosmological scales, providing the seeds for the structure of the universe \cite{Guth:1980zm, Linde:1981mu, Albrecht:1982wi}. Couplings between the inflaton and gauge fields can amplify gauge-field fluctuations to large amplitudes, sourcing primordial magnetic fields and gravitational waves that carry imprints of inflationary dynamics \cite{Durrer:2013pga, Subramanian:2015lua, Vachaspati:2020blt}. Strong electric fields produced in this process generate charged particles through the Schwinger effect \cite{Sauter:1931zz, Heisenberg:1936nmg, Schwinger:1951nm, Kluger:1991ib}, giving rise to a conducting plasma. Thus, gauge-field production and magnetogenesis are intrinsically connected to the formation and dynamics of the plasma in which the fields evolve.

The subsequent evolution of primordial magnetic fields is commonly described within magnetohydrodynamics (MHD), where the magnetic field is evolved from prescribed initial conditions. These initial conditions are typically chosen independently of the microscopic mechanism that generated the fields \cite{Christensson:2000sp, Banerjee:2004df, Reppin+Banerjee17, Trivedi+18, Bhat+21, HS23, Isochrone26}, and the plasma is introduced as an already conducting medium with specified properties, or under simplified prescriptions \cite{BHKRS21, BS21, BHS21, BP23, Gorbar:2023zla}. Such an approach captures the later stages of turbulent magnetic evolution, but leaves open how the plasma and its bulk motions emerge from magnetogenesis itself. Establishing this connection is essential for robust predictions of primordial magnetic fields and gravitational waves, and for distinguishing their possible origins with upcoming observations.

In this Letter we bridge this gap by presenting, for the first time, the coupled inhomogeneous inflaton--plasma equations and numerically following their evolution continuously from inflation through reheating and into the MHD regime. Rather than imposing MHD initial conditions, we dynamically generate the conducting plasma and its inhomogeneities and evolve their nonlinear interaction with the gauge fields. This self-consistent evolution reveals a previously unexplored turbulent regime driven directly by the inflaton dynamics.

We demonstrate our formalism using axion inflation \cite{Freese:1990rb, Svrcek:2006yi, Anber:2009ua, Barnaby:2011qe}, while it can be applied more broadly. The pseudoscalar axion field $\phi$, which acts as the inflaton, couples to the electromagnetic field through the Chern--Simons interaction
$\frac{\alpha}{4f}\phi F_{\mu\nu}\tilde{F}^{\mu\nu}$, leading to exponential gauge-field amplification \cite{Anber:2009ua, Barnaby:2011qe, Pajer:2013fsa}. 
Axion inflation provides a well-motivated framework for primordial magnetogenesis \cite{Garretson:1992vt, Anber:2006xt, Fujita:2015iga, Adshead:2016iae, Durrer:2023rhc} and gravitational-wave production \cite{Sorbo:2011rz, Cook:2013xea, Domcke:2016bkh, Garcia-Bellido:2016dkw, Bastero-Gil:2022fme, Garcia-Bellido:2023ser, Corba:2024tfz, Corba:2025reo, vonEckardstein:2025oic, vonEckardstein:2025elq} and has recently been explored with lattice simulations \cite{Adshead:2019igv, Adshead:2019lbr, Adshead:2023mvt, Caravano:2021bfn, Caravano:2022epk, Figueroa:2023oxc, Caravano:2024xsb, Sharma:2024nfu, Figueroa:2024rkr, Iarygina:2025ncl}.

We show that inflaton dynamics drive an initial turbulent phase, which we call \textit{inflaton-driven turbulence}.  We further find that kinetic energy dominates over magnetic energy, driving dynamo amplification and indicating strong plasma motions. The inflaton-driven turbulence leads to a steeper magnetic energy decay than magnetically dominated helical turbulence.
These results show that the plasma dynamics generated during magnetogenesis can alter the subsequent evolution of primordial magnetic fields, motivating self-consistent non-perturbative treatments when connecting primordial fields to present-day observables.

\noindent\textbf{From inflation to turbulence.}\\
It is convenient to work with 
comoving electric and magnetic fields, defined as
$\bm{E}=-\partial_\tau \bm{A}+\bm{\nabla}A_0$, $\bm{B}=\bm{\nabla}\times \bm{A}$, where $A_\mu=(A_0,\bm{A})$ is the Abelian vector potential and we use derivatives with respect to conformal time $d\tau=dt/a(t)$, with $a(t)$ being the scale factor.
We do not rescale the axion field,
$\phi=\phi^{\rm{ph}}$, to retain the standard equation of motion and avoid the additional term involving the second conformal-time derivative of the scale factor.

\begin{figure*}[t!]
\centering
\includegraphics[width=\textwidth]{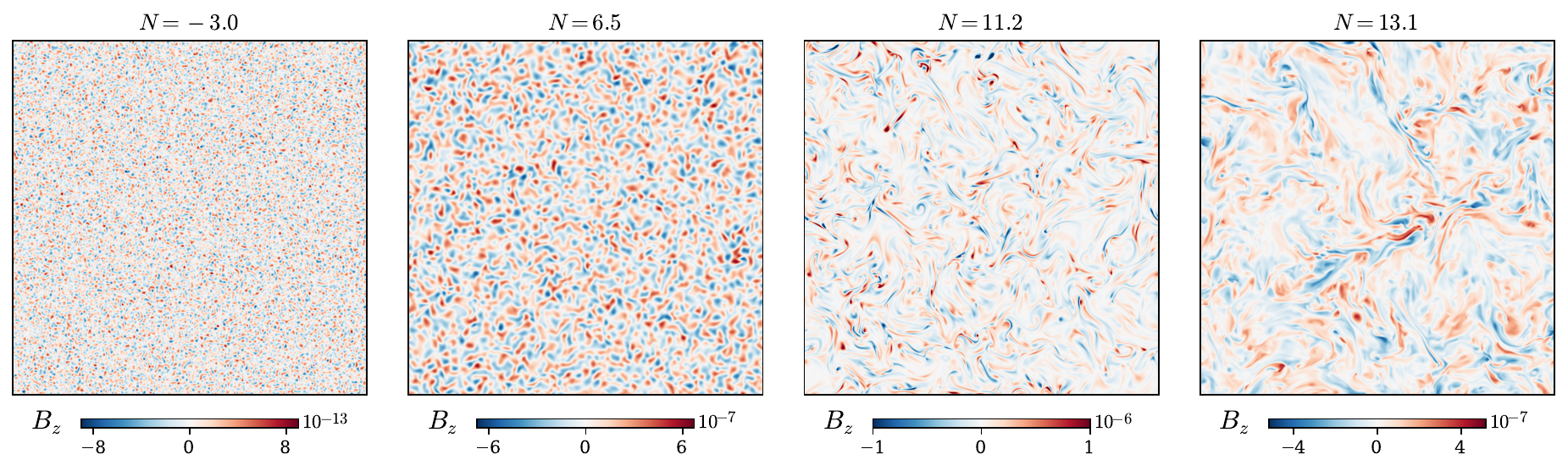}
\caption{Snapshots of the magnetic field evolution, starting from the Bunch--Davies vacuum through the onset of turbulence and subsequent turbulent decay. The magnetic perturbations gradually grow and, with the onset of turbulence, develop increasingly twisted and coherent structures. $N$ denotes the number of e-folds, and the color indicates the amplitude of the fluctuations.}
%3D/Hbdn512alpf90_rho28_ampl1_Gam9o2
\label{fig:snapshots}
\end{figure*}

We begin by introducing the equations of motion during inflation. In addition 
to the gauge-field dynamics, we include bulk plasma motions and fully 
inhomogeneous energy densities, allowing us to consistently describe the 
backreaction of the produced fields and the subsequent magnetohydrodynamic 
evolution. The equations of motion are 
\begin{align}
&\partial_\tau^2\phi+2{\cal{H}}\partial_\tau\phi-\bm{\nabla}^2\phi +a^2\frac{dV}{d \phi} =\frac{\alpha}{ a^2 f}\bm{E}\cdot\bm{B}, \label{eq:phi}\\
& \partial_\tau\bm{E}-{\rm{rot}}\, \bm{B}+\frac{\alpha}{f}\left(\partial_\tau\phi\bm{B}+\bm{\nabla}\phi\times \bm{E}\right)+\bm{J}=0,\label{eq:Edotconf}\\
&\bm{\nabla}\cdot \bm{E}=-\frac{\alpha}{ f}\bm{\nabla}\phi\cdot \bm{B}, \quad \bm{\nabla}\cdot \bm{B}=0, \label{eq:nablaE}\\
&\partial_\tau{\bm{B}}+{\rm{rot}}\,\bm{E}=0,\\
&{\cal H}^2=\frac{8\pi}{3m_{\rm {Pl}}^2 a^2}\langle\rho_{\phi}+\rho_{E}+\rho_{B}+\rho\rangle, \label{eq:H}
\end{align}
where ${\cal H}=\partial_\tau a/a$ is the conformal Hubble parameter, $V(\phi)$ is the axion potential, $\alpha$ is the axion--gauge coupling, $f$ is the axion decay constant and $\bm{J}$ is
the electric current density, which includes the contribution from bulk plasma motions and is given by
\begin{equation} \label{eq:J}
  \bm{J} = \sigma_E \left(\bm{E}+\bm{u}\times\bm{B}\right).
\end{equation}
Here, $\bm{u}$ is the bulk velocity of the
plasma and $\sigma_E$ is the electric conductivity \footnote{In general, the current can be also written as $\bm{J} = \sigma_E (\bm{E}+\bm{u}\times\bm{B}) + \sigma_B \bm{B}$. Since $\sigma_B \bm{B}$ contribution is subdominant, see \cite{vonEckardstein:2024tix, Iarygina:2025ncl}, we omit it here. The contribution $\bm{u}\times\bm{B}$ is absent under the assumption of plasma homogeneity \cite{Gorbar:2023zla}.}. The comoving energy densities are  
$\rho_{\phi}= a^2(\partial_\tau \phi)^2/2 +a^2(\bm{\nabla}\phi)^2/2 +a^4V $  for the axion,
$\rho_E= \bm{E}^2 /2$ for the electric field, $\rho_B=\bm{B}^2 /2$
for the magnetic field, and $\rho$ is the radiation energy density. We also define the kinetic energy density of the plasma, $\rho_\rm{kin}=\textstyle{\frac{1}{2 }} \left(\rho+p\right) \bm{u}^2 $, where $p$ is the pressure of the plasma \footnote{In our earlier work \cite{Iarygina:2025ncl}, the symbol $\rho_\chi$ denoted the 
energy density transferred to fermions (and the resulting plasma) through electromagnetic dissipation. However, part of this energy is 
first transferred into kinetic energy through the work done by the Lorentz force, $\langle\bm{u}\cdot(\bm{J}\times\bm{B})\rangle$, before being dissipated through viscous effects,  while the remaining part is directly converted into radiation through resistive 
heating. Therefore, $\rho_\chi=\rho+\rho_{\rm{kin}}$ represents the total energy transferred from the electromagnetic fields to the plasma.}. The comoving energy densities are related to the physical ones via $\rho_i=a^4\rho_i^\rm{ph}$, while the comoving and physical bulk velocities of the plasma are the same, $ \bm{u}= \bm{u}^\rm{ph}$. Angle brackets $\langle ...\rangle$ denote volume averaging over the whole simulation domain.

We now turn to the equations describing the turbulent plasma evolution. To establish a self-consistent connection with MHD, we treat the energy density fully inhomogeneously, allowing for spatial variations generated by plasma motions and electromagnetic interactions. The radiation energy density $\rho$ evolves according to
\begin{equation}
   \partial_\tau \rho+\bm{\nabla}\cdot\left(\left(\rho+p\right)\bm{u}-\kappa\bm{\nabla}\rho\right)
=\bm{J}\cdot \bm{E}+2\rho\nu\bm{\mathsf{S}}^2, \label{eq:rho}
\end{equation}
where the first term on the right-hand side describes the transfer of electromagnetic
energy into the plasma through resistive heating, while the second term accounts for
viscous dissipation of kinetic energy. The parameter $\kappa$ denotes the thermal
diffusivity, $\nu$ is the kinematic viscosity, and $\bm{\mathsf{S}}$ is the rate-of-strain
tensor describing the local velocity gradients with the components
$\mathsf{S}_{ij}=(\partial_i u_j+\partial_j u_i)/2-\delta_{ij}\partial_k u_k/3$.

An additional important component of the MHD evolution is the plasma velocity. 
Magnetic fields exert a Lorentz force, $\bm{J}\times\bm{B}$, on the charged plasma, generating bulk 
motions that can become turbulent. These velocity fields, in turn, modify the 
evolution of the electromagnetic fields through the $\bm{u}\times\bm{B}$ 
contribution to the current density, requiring a self-consistent treatment of 
the coupled plasma-field dynamics. The bulk velocity of the plasma obeys the 
momentum equation
\begin{equation}
    \left(\rho+p\right)\left(\partial_\tau\bm{u}+\bm{u}\cdot\bm{\nabla}\bm{u}\right)
+\left(\bm{\nabla}p+\bm{u}\,\partial_\tau p\right)
=\bm{J}\times\bm{B}+\bm{\nabla}\cdot(2\rho\nu\bm{\mathsf{S}}). \label{eq:u}
\end{equation}
Here the left-hand side describes the evolution of the plasma momentum, including
inertial and pressure-gradient contributions. The Lorentz force provides the coupling between the electromagnetic fields
and the plasma motion, while the final term accounts for viscous momentum
transport. Together with the radiation energy equation, this provides a
self-consistent description of the turbulent MHD evolution of the primordial
plasma.\\

\noindent\textbf{Numerical implementation.}\\
We perform numerical simulations on a lattice using the {\sc Pencil Code}~\cite{PencilCode:2020eyn} with a grid resolution of $512^3$ points.
The evolution is initialized approximately $3$ $e$-folds before the end of inflation, with the fields set to the Bunch--Davies initial conditions.
The axion potential is taken to be quadratic,
$V(\phi)=\frac{1}{2}m_{\phi}^2\phi^2$,
which provides a good description of the dynamics close to the end of inflation. We use $m_{\phi}= 10^{-6}\, m_{\rm {Pl}}$ and $\alpha m_{\rm {Pl}}/f=90$.

For the electric conductivity we set \cite{Bavarsad:2017oyv, Domcke:2018eki, Domcke:2019qmm, vonEckardstein:2024tix}
\begin{equation}
    \sigma_E = \frac{(e |Q|)^3}{6\pi^2 \mathcal{H}} B \coth\left( \frac{\pi B}{E} \right), \label{eq:sigmaE} 
\end{equation}
where $E=|\bm{E}|$ is the magnitude of the electric field and $B$ is the magnetic field, projected onto the direction of the electric field.
Several descriptions of the conductivity have been studied in the literature \cite{Bavarsad:2017oyv, Domcke:2018eki, Sobol:2019xls, Domcke:2019qmm, Gorbar:2021rlt, Fujita:2022fwc, vonEckardstein:2024tix},
but here we employ a collinear description (see \cite{Iarygina:2025ncl, vonEckardstein:2024tix}) where $E$ and $B$ are the comoving rms electric and magnetic fields.
Above, $e$ is the gauge coupling constant, $Q$ is the particle’s charge with $Q^3=41/12$. In our simulations we use the running gauge coupling defined as 
    $e\equiv g'(\tilde{\mu})=\left([g'(m_Z)]^{-2}+\frac{41}{48 \pi^2}\ln\frac{m_Z}{\tilde{\mu}} \right)^{-1/2}$,
where $g'(m_Z)\simeq 0.35$, $m_Z\simeq 91.2 \, \rm{GeV}$ with $ \tilde{\mu}=(\rho_E+\rho_B)^{1/4}$.
We consider the strong-field regime, $|eQE|\gg {\cal H}^2$, and take the produced plasma to be effectively massless.

For a radiation-dominated plasma, the equation of state is $p=\rho/3$, where the speed of light is set to unity. Therefore, the enthalpy density entering the fluid equations is
$\rho+p=\frac{4}{3}\rho$. We use this relation throughout the simulations.

When particle production via the Schwinger effect is consistently included, lattice simulations~\cite{Iarygina:2025ncl} show that gauge preheating does not occur even for large axion-gauge couplings. Therefore, the transition to the radiation-dominated era has to proceed through perturbative inflaton decay. To account for this process, we include the axion decay width into two photons \cite{ParticleDataGroup:2014cgo, Adshead:2015pva} 
\begin{equation}
    \Gamma_{\phi0}=\frac{\alpha^2 m_{\phi}^3}{64\pi f^2},
\label{Gamma_phi}
\end{equation}
which enters the axion and radiation energy equations as an additional
dissipative term. This introduces a friction term in the axion equation,
driving the decay of the inflaton, while simultaneously providing a source
term for the radiation energy density
\begin{align}
&\partial_\tau^2\phi+(2{\cal{H}}+a\Gamma_\phi)\partial_\tau\phi-\bm{\nabla}^2\phi +a^2\frac{dV}{d \phi} =\frac{\alpha}{ a^2 f}\bm{E}\cdot\bm{B},  \label{eq:phirhoGamma}\\
 &\partial_\tau \rho+\bm{\nabla}\cdot\left({\textstyle\frac{4}{3}}\rho\bm{u}-\kappa\bm{\nabla}\rho\right)
=a\Gamma_\phi\rho_\phi+\bm{J}\cdot \bm{E}+2\rho\nu\bm{\mathsf{S}}^2.\notag
\end{align}

The first term on the right-hand side of the radiation equation describes energy transfer from the decaying axion field, while the remaining terms account for electromagnetic dissipation and viscous heating in the plasma.

To avoid an instantaneous transition to the radiation-dominated era, we turn
on the inflaton decay smoothly around the end of inflation,
\begin{equation}
    \Gamma_\phi(N)=\Gamma_{\phi0}\Theta_{\Delta N}(N-N_{\rm{end}}), \label{eq:Gamma}
\end{equation}
where $N$ is the number of $e$-folds after the end of inflation and $\Theta_{\Delta N}$ is a smooth step function with transition width
$\Delta N$ in $e$-folds. This provides a gradual transfer of energy from the
inflaton to the radiation bath. For the results presented here we employ $\Gamma_{\phi0}=10^{-9}m_{\rm {Pl}}$. Larger values of $\Gamma_{\phi0}$ would correspond to a shorter reheating period, while smaller values would result in a longer one.

In MHD, viscosity and magnetic diffusivity play analogous roles: the kinematic viscosity $\nu$ smooths gradients in the velocity field, while the magnetic diffusivity $ 1/\sigma_E$ smooths magnetic-field gradients. Their ratio defines the magnetic Prandtl number,
\begin{equation}
    \rm{Pr}_{\rm{M}}=\nu\sigma_E. 
\end{equation}
The value of $\rm{Pr}_{\rm{M}}$ determines the relative efficiency of viscous and magnetic diffusion and therefore controls the separation between the kinetic and magnetic dissipation scales.
In numerical simulations we treat $\rm{Pr}_{\rm{M}}$ as a parameter and investigate the values $ \rm{Pr}_{\rm{M}}=1,10$. This range allows us to assess the sensitivity of the magnetic-field evolution and turbulent dynamics to the relative strength of viscous and magnetic dissipation. We also fix the thermal diffusivity $\kappa=2 \times 10^4\,m_{\rm {Pl}}^{-1}$.

We use the resistive gauge \cite{Candelaresi:2010im}
\begin{equation}\label{eq:ResGauge}
A_0=\sigma_E^{-1}\nabla\cdot\bm{A}, 
\end{equation}
which is particularly convenient for simulations of highly conducting plasmas.
In the limit of large conductivity, the displacement current can be neglected,
and the Maxwell equations reduce to the resistive induction equation. In this
gauge, the uncurled form of the induction equation contains an explicit
diffusion term proportional to $\nabla^2\bm{A}$, making the numerical treatment
of magnetic-field evolution more stable and efficient.  We employ this gauge
from the beginning of the simulations.

\begin{figure}[t!]\begin{center}
\includegraphics[width=\columnwidth]{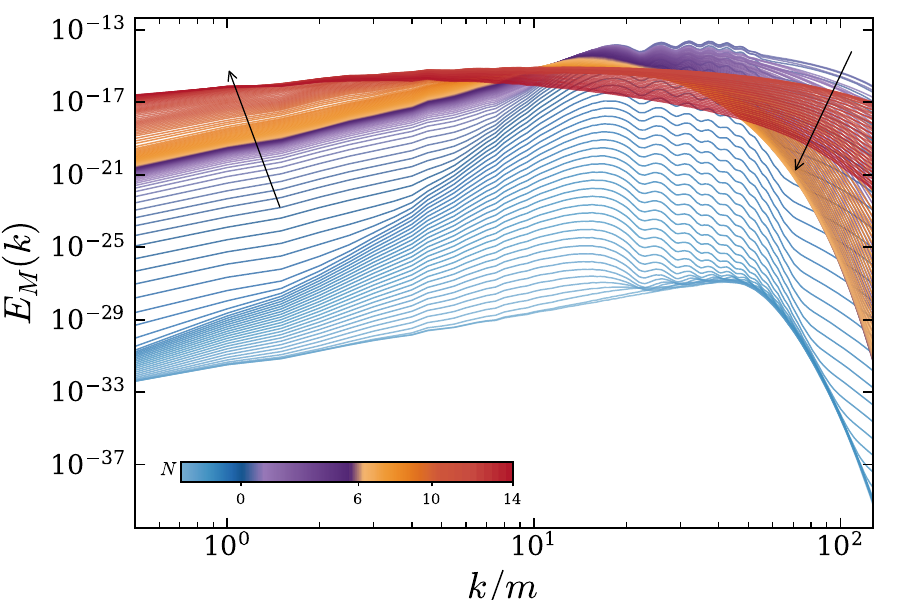}\\
\includegraphics[width=\columnwidth]{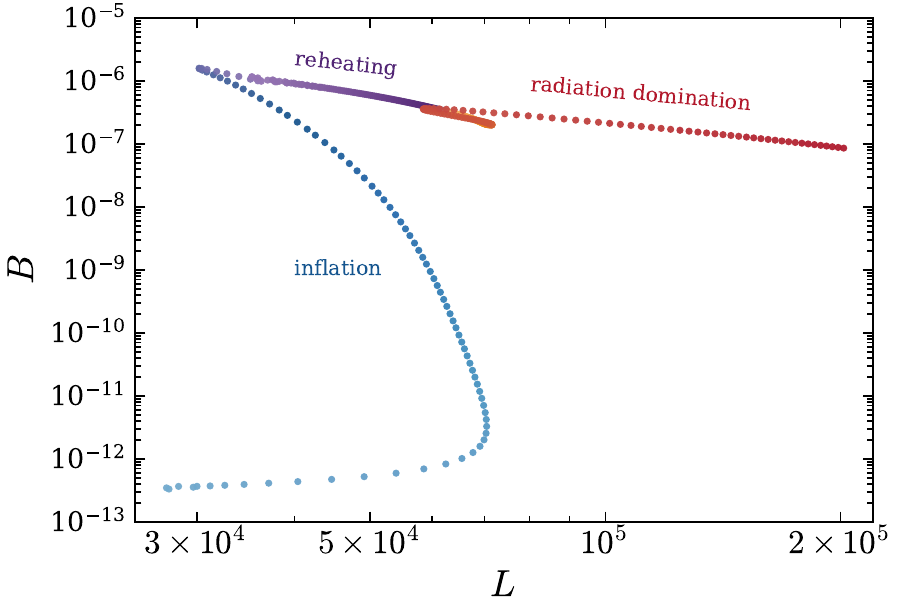}
%Hbdn512alpf90_rho28_ampl1_Gam9o2 
\end{center}\caption[]{\textit{Top panel:} Magnetic energy spectra showing magnetic field growth during inflation (light to dark blue), followed by an inflaton-driven turbulent regime during reheating (light to dark violet), with decreasing magnetic energy and increasing coherence length due to an inverse cascade. This is followed by decaying turbulence (light to dark orange) and renewed magnetic-field growth driven by dynamo action (red) with a subsequent decay during radiation domination. Colors correspond to the number of $e$-folds displayed in the color bar. Animations of the power-spectrum evolution and magnetic-field snapshots from Figure \ref{fig:snapshots} are available at this \href{https://github.com/OksanaIarygina/Turbulence-from-inflation}{link}. \textit{Bottom panel:} Evolution of magnetic field amplitude versus the coherent length for the same color coding, in Planck units.
}\label{fig:Power}\end{figure}

\begin{figure}[t!]\begin{center}
\includegraphics[width=\columnwidth]{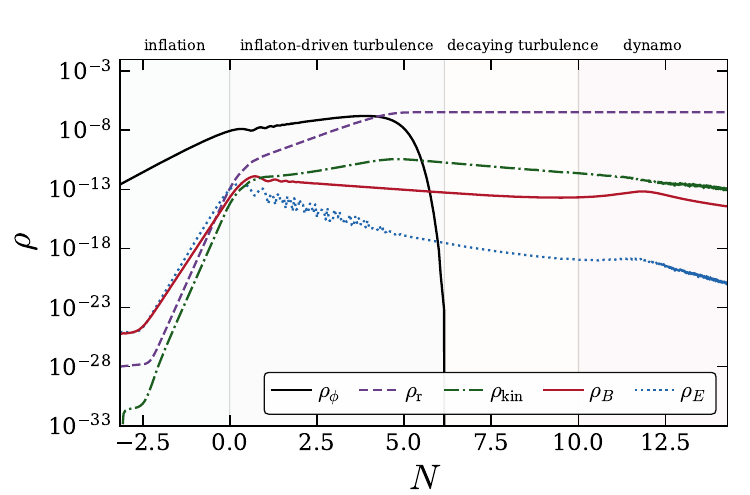}
%Hbdn512alpf90_rho28_ampl1_Gam9o2  
\end{center}\caption[]{Evolution of the energy densities of the axion $\rho_{\phi}$ (solid black), radiation $\rho_{\rm r}$ (dashed violet), kinetic energy $\rho_{\rm{kin}}$ (dash-dotted green), magnetic energy $\rho_B$ (solid red), and electric energy $\rho_E$ (dotted green), as a function of the number of $e$-folds $N$. Vertical grid lines mark the boundaries between the different regimes: inflation (blue-shaded area), axion-driven turbulence (violet-shaded area), decaying turbulence (orange-shaded area), and the onset of dynamo action (red-shaded area) marked by magnetic field amplification with a characteristic bump, followed by subsequent decay. Starting from the onset of reheating, the kinetic energy dominates over the magnetic energy.
}\label{fig:energies}\end{figure}

After reheating, the produced plasma becomes highly conducting, and the
system enters the MHD regime. In this limit, the
displacement current, $\partial_\tau\bm{E}$, in equation \eqref{eq:Edotconf} can be neglected.
Physically, this corresponds to the rapid damping of electric fields due to the
large conductivity of the plasma, such that electromagnetic fields adjust on
timescales much shorter than the characteristic timescale of plasma motions.
The electromagnetic evolution is therefore described by the resistive induction
equation, while the electric field is obtained algebraically from the generalized
Ohm's law rather than being evolved as an independent dynamical variable.
%This approximation removes the need to resolve the fast electromagnetic propagation
This approximation is applied for $N>2$, and we have verified that the results are insensitive to choosing a later value of $N$. It removes the need to resolve the fast electromagnetic propagation
timescale and allows us to focus on the slower turbulent evolution of the
primordial plasma. Hence, we numerically solve \eqref{eq:Edotconf}--\eqref{eq:H}, \eqref{eq:u}, \eqref{eq:phirhoGamma} with \eqref{eq:J}, \eqref{eq:sigmaE} and \eqref{eq:ResGauge}.\\

\noindent\textbf{Results.}\\
Figure \ref{fig:snapshots} shows snapshots of the magnetic-field evolution as a function of the number of $e$-folds. Initially, the field consists of Gaussian random fluctuations corresponding to the Bunch--Davies vacuum. As the inflaton evolves, these fluctuations are amplified, leading to the growth of the magnetic field and the development of increasingly pronounced spatial structure. The subsequent evolution is characterized by the onset of turbulence driven by the inflaton dynamics, followed by turbulent decay. Brighter colors indicate stronger magnetic fields, while the evolving spatial structure reflects the transfer of magnetic energy across scales throughout these stages.

The onset of turbulence is demonstrated by the evolution of the magnetic energy spectrum, defined as 
\begin{equation}
E_M(k,\tau)={\textstyle\frac{1}{2}}\,\mathrm{Sp}\!\left[\bm{B}(\bm{k},\tau)\right],
\;\;
{\textstyle\int} E_M(k,\tau)\,dk=\langle\rho_B\rangle ,
\end{equation}
which characterizes the distribution of magnetic energy over comoving
wavenumbers. In the top panel of Figure~\ref{fig:Power}, we show the magnetic energy spectra at different stages of the evolution.
During inflation, the magnetic energy spectrum increases due to axion-driven gauge field production (blue to dark blue).
After the end of inflation and during reheating, the peak of the spectrum moves toward smaller wavenumbers, corresponding to larger length scales,
while the overall power decreases, indicating the onset of the inverse cascade (light to dark violet). In this regime, the inflaton energy transfer and electromagnetic dissipation balance each other, as demonstrated in the Supplemental Material \footnote{In the Supplemental Material, which includes Ref.~\cite{Jones08}, we show the electromagnetic energy balance and demonstrate that the initial inverse-cascade phase is driven by the inflaton energy-transfer term. We further discuss the dynamo action and show that the magnetic energy bump is absent in 2D, where dynamo action is not possible.}, establishing a distinct turbulent phase that we call inflaton-driven turbulence.
Afterward, the system enters a turbulent decay phase (orange), with a temporary increase in magnetic energy at both large and small scales, signaling the onset of dynamo amplification (red).
Dynamo action is the result of a magnetic instability that transfers kinetic energy to magnetic energy.
Even for decaying turbulence, it can lead to an episode of nearly exponential growth before the magnetic field reaches a certain fraction of the kinetic energy \cite{Bra+19,BN25}.
It requires that the work done by the Lorentz force is negative and in excess of the electromagnetic energy dissipation, which is here the case, see \footnotemark[3].
The colors in the color bar correspond to the different stages of the background evolution (inflation, reheating, inflaton-driven inverse cascade, turbulent decay, dynamo), which we also discuss in more detail below in Figure~\ref{fig:energies}.

The bottom panel of Figure~\ref{fig:Power} shows the corresponding evolution of the comoving magnetic field as a function of the comoving coherence length, following the same color-coding as the top panel. 
The coherence length is defined as
\begin{equation}
L=\frac{\int E_M(k,\tau)\,k^{-1}\,dk}{\int E_M(k,\tau)\,dk}.
\end{equation}
We see the magnetic field grows from its initial vacuum fluctuations
through inflation and up to the onset of reheating. During the subsequent inflaton-driven turbulent regime, the magnetic field decays while the coherence length increases, characteristic of an inverse cascade. This behavior continues into the subsequent turbulent decay regime, with the field strength decreasing as magnetic energy is transferred toward larger scales. With the onset of dynamo action, the magnetic field undergoes renewed amplification, temporarily reversing the decay of the field. At later times, the field resumes its decay, reflecting the transition to a turbulent decay regime. We find that the decay follows a steeper scaling, $B\sim L^{-1}$, rather than conventional $B\sim L^{-1/2}$ behavior expected for fully developed helical turbulence
\cite{Hatori84,BM99,BK17}.
As the plasma evolves toward a more highly conducting regime, we expect the decay to gradually approach the latter scaling. The initial $B\sim L^{-1}$ regime therefore represents an important transient stage, lasting for $\sim14$ $e$-folds, that should be taken into account when predicting the magnetic field amplitude and coherence length at the present epoch.

The transition between the different regimes is further illustrated by the evolution of the energy densities, which traces the energy transfer between the inflaton, electromagnetic fields, and plasma, as shown in Figure~\ref{fig:energies}. The inflaton energy density (solid black) dominates until the end of reheating, when the inflaton decays and transfers its energy to the radiation bath. Throughout inflation, the magnetic and electric energy densities (solid red and dotted blue) grow, accompanied by the gradual production of plasma through the Schwinger effect, reflected in the radiation energy density (dashed violet). The magnetic field drives plasma motions through the Lorentz force, leading to the growth of the kinetic energy density (dash-dotted green). A substantial fraction of the inflaton energy is transferred to plasma kinetic energy during reheating, particularly during inflaton decay.
Crucially, it even exceeds the magnetic energy, contrary to the commonly assumed picture in which the kinetic energy is subdominant
\cite{Bra+17,Bra+19}.
The magnetic-field evolution transitions from a magnetically dominated regime to one controlled by plasma motions, enabling dynamo amplification and producing the characteristic bump in the magnetic energy \footnotemark[3]. \\

\noindent\textbf{Conclusions and outlook.}\\
We conducted the first non-perturbative study, analytically formulating the coupled dynamics and numerically following the nonlinear evolution of primordial fields from vacuum fluctuations during inflation to the emergence of turbulence.
Our findings revealed an initial turbulent phase driven by the inflaton dynamics, which we term \textit{inflaton-driven turbulence}. This nonlinear evolution results from the electromagnetic energy transfer from the inflaton to the gauge fields and their subsequent interaction with the generated plasma, establishing that the inflaton both can seed the primordial magnetic field and initiate its subsequent nonlinear magnetohydrodynamic evolution. We found that the initial turbulent phase produces a steeper decay than expected for magnetically-dominated helical turbulence, which must be accounted for in determining the present-day properties of primordial magnetic fields.

Additionally, we found that kinetic energy dominates over magnetic energy starting from the onset of turbulence.
Following axion decay, the system enters a kinetically dominated turbulent regime driven by strong bulk plasma motions, which drive dynamo amplification of the magnetic field. The strong plasma motions suggest a possible connection to the steeper magnetic-field decay.
This highlights the importance of bulk motions for the nonlinear evolution of primordial magnetic fields and calls for their inclusion in MHD treatments of primordial magnetogenesis.

Our analysis focused on axion inflation, but the framework naturally extends to other mechanisms of gauge-field generation during inflation and their subsequent magnetohydrodynamic evolution.

These results have important implications for probing the early universe through gravitational waves and primordial magnetic fields. The strong gauge field instabilities can source gravitational waves, followed by an additional contribution from the turbulent bulk motion of the generated plasma \cite{Caprini+Durrer01, Kahniashvili+05, Caprini+Durrer06, Kahniashvili+08, RoperPol:2019wvy}. In our framework, these contributions emerge from the same coupled gauge field and plasma evolution: the first is associated with the growth and peak of the gauge fields, while the second arises from the subsequent turbulent plasma dynamics. This connection can imprint characteristic scale dependence on the gravitational wave spectrum, providing a distinctive signature that can be probed by observatories such as Pulsar Timing Arrays \cite{NANOGrav:2023gor}, LISA \cite{LISA:2017pwj}, and the Einstein Telescope \cite{ET:2025xjr}.
Direct lattice simulations of the magnetic field evolution can further determine its decay laws, amplitude, and coherence length, enabling quantitative comparisons with blazar observations \cite{Neronov:2010gir, MAGIC:2022piy, Blunier+26}, and future measurements with CTAO \cite{CTAConsortium:2017dvg},  LHAASO \cite{LHAASO:2019qtb}, SKA \cite{Weltman:2018zrl}, LOFAR \cite{LOFAR:2013jil}. Combining these measurements with gravitational-wave observations can help distinguish inflationary magnetogenesis from primordial magnetic fields generated during phase transitions in the early universe \cite{Kahniashvili:2009mf, Caprini:2009yp, Neronov+21, Bra+21}. These results also provide a non-perturbative window into inflaton dynamics, capturing the growth of inhomogeneities and energy transfer to the gauge fields, while allowing for future studies of inflaton fragmentation and the subsequent thermalization of the generated plasma \cite{ Podolsky:2005bw, Felder:2006cc}. Extending this treatment to different inflaton masses, couplings, and particle contents can establish how the microscopic physics drives the generation and evolution the primordial plasma, magnetic fields, and gravitational waves, providing a direct connection between inflationary particle physics and observable signatures in the late universe.

More broadly, the magnetic field spectra, amplitudes, and coherence lengths obtained from direct numerical simulations can provide physically motivated initial conditions for MHD studies of the early universe plasma and astrophysical systems \cite{Kahniashvili:2012uj, Kahniashvili:2015msa, Bhat+21, HS23, Isochrone26}. In particular, these results can be used to model the evolution of primordial magnetic fields through recombination, where the resulting plasma inhomogeneities can modify the recombination history and the inferred sound horizon, providing a direct connection to the Hubble tension \cite{JP20, Paoletti+22, Jedamzik:2023rfd, Jedamzik:2025cax, Schiff+Venumadhav25}. The same initial conditions can be propagated into simulations of large-scale structure formation \cite{Mtchedlidze:2021bfy},
offering a unified framework for connecting inflationary magnetogenesis to magnetic fields observed across astrophysical scales
\cite{Vazza+17, Tjemsland+24, Neronov+24}.

\noindent\textbf{Data availability.}
The source code used for the simulations of this study, the {\sc Pencil Code},
is freely available from Refs.~\cite{PencilCode:2020eyn,PC}.
The simulation setups and the corresponding data are freely available from 
Ref.~\cite{DATA}, and animations are available via this \href{https://github.com/OksanaIarygina/Turbulence-from-inflation}{link}.
\\

\noindent\textbf{Acknowledgements.} We thank Evangelos I.\ Sfakianakis for useful comments on the draft and initial discussions on pair plasmas. The work of OI is supported by the Swedish Research Council (Vetenskapsr{\aa}det) under the Starting Grant No.\ 2025-04140.
AB acknowledges support from the European Research Council through the ERC Synergy Grant COSMOMAG under grant No.\ 101224803,
the Swedish Research Council (Vetenskapsr{\aa}det) under grant No.\ 2025-05957,
the National Science Foundation under grant Nos.\ NSF AST-2307698, AST-2408411, and NASA Award 80NSSC22K0825.
We thank the Swedish National Allocations Committee for providing computing resources at the Center for Parallel Computers
at the Royal Institute of Technology in Stockholm and the National Supercomputer Centre (NSC) at Link\"oping. 
\appendix

\bibliography{refs}

@article{Fujita:2022fwc,
    author = "Fujita, Tomohiro and Kume, Jun'ya and Mukaida, Kyohei and Tada, Yuichiro",
    title = "{Effective treatment of U(1) gauge field and charged particles in axion inflation}",
    eprint = "2204.01180",
    archivePrefix = "arXiv",
    primaryClass = "hep-ph",
    reportNumber = "RESCEU-3/22, KEK-TH-2402",
    doi = "10.1088/1475-7516/2022/09/023",
    journal = "JCAP",
    volume = "09",
    pages = "023",
    year = "2022"
}

@article{Figueroa:2023oxc,
    author = "Figueroa, Daniel G. and Lizarraga, Joanes and Urio, Ander and Urrestilla, Jon",
    title = "{Strong Backreaction Regime in Axion Inflation}",
    eprint = "2303.17436",
    archivePrefix = "arXiv",
    primaryClass = "astro-ph.CO",
    doi = "10.1103/PhysRevLett.131.151003",
    journal = "Phys. Rev. Lett.",
    volume = "131",
    number = "15",
    pages = "151003",
    year = "2023"
}

@article{Sobol:2019xls,
    author = "Sobol, O. O. and Gorbar, E. V. and Vilchinskii, S. I.",
    title = "{Backreaction of electromagnetic fields and the Schwinger effect in pseudoscalar inflation magnetogenesis}",
    eprint = "1907.10443",
    archivePrefix = "arXiv",
    primaryClass = "astro-ph.CO",
    doi = "10.1103/PhysRevD.100.063523",
    journal = "Phys. Rev. D",
    volume = "100",
    number = "6",
    pages = "063523",
    year = "2019"
}

@article{Sharma:2024nfu,
    author = "Sharma, Ramkishor and Brandenburg, Axel and Subramanian, Kandaswamy and Vikman, Alexander",
    title = "{Lattice simulations of axion-U(1) inflation: gravitational waves, magnetic fields, and scalar statistics}",
    eprint = "2411.04854",
    archivePrefix = "arXiv",
    primaryClass = "astro-ph.CO",
    reportNumber = "NORDITA-2024-040",
    doi = "10.1088/1475-7516/2025/05/079",
    journal = "JCAP",
    volume = "05",
    pages = "079",
    year = "2025"
}

@ARTICLE{BP23,
       author = {{Brandenburg}, Axel and {Protiti}, Nousaba Nasrin},
        title = "{Electromagnetic Conversion into Kinetic and Thermal Energies}",
      journal = {Entropy},
         year = 2023,
        month = aug,
       volume = {25},
       number = {9},
          eid = {1270},
        pages = {1270},
          doi = {10.3390/e25091270},
archivePrefix = {arXiv},
       eprint = {2308.00662},
 primaryClass = {physics.plasm-ph},
       adsurl = {https://ui.adsabs.harvard.edu/abs/2023Entrp..25.1270B}
}

@article{Bavarsad:2017oyv,
    author = "Bavarsad, Ehsan and Kim, Sang Pyo and Stahl, Cl\'ement and Xue, She-Sheng",
    title = "{Effect of a magnetic field on Schwinger mechanism in de Sitter spacetime}",
    eprint = "1707.03975",
    archivePrefix = "arXiv",
    primaryClass = "hep-th",
    doi = "10.1103/PhysRevD.97.025017",
    journal = "Phys. Rev. D",
    volume = "97",
    number = "2",
    pages = "025017",
    year = "2018"
}

@article{Domcke:2018eki,
    author = "Domcke, Valerie and Mukaida, Kyohei",
    title = "{Gauge Field and Fermion Production during Axion Inflation}",
    eprint = "1806.08769",
    archivePrefix = "arXiv",
    primaryClass = "hep-ph",
    reportNumber = "DESY 18-098, DESY-18-098",
    doi = "10.1088/1475-7516/2018/11/020",
    journal = "JCAP",
    volume = "11",
    pages = "020",
    year = "2018"
}

@article{Domcke:2019qmm,
    author = "Domcke, Valerie and Ema, Yohei and Mukaida, Kyohei",
    title = "{Chiral Anomaly, Schwinger Effect, Euler-Heisenberg Lagrangian, and application to axion inflation}",
    eprint = "1910.01205",
    archivePrefix = "arXiv",
    primaryClass = "hep-ph",
    reportNumber = "DESY-19-166, DESY 19-166",
    doi = "10.1007/JHEP02(2020)055",
    journal = "JHEP",
    volume = "02",
    pages = "055",
    year = "2020"
}

@article{PencilCode:2020eyn,
       author = {{Pencil Code Collaboration} and {Brandenburg}, Axel and {Johansen}, Anders and {Bourdin}, Philippe and {Dobler}, Wolfgang and {Lyra}, Wladimir and {Rheinhardt}, Matthias and {Bingert}, Sven and {Haugen}, Nils and {Mee}, Antony and {Gent}, Frederick and {Babkovskaia}, Natalia and {Yang}, Chao-Chin and {Heinemann}, Tobias and {Dintrans}, Boris and {Mitra}, Dhrubaditya and {Candelaresi}, Simon and {Warnecke}, J{\"o}rn and {K{\"a}pyl{\"a}}, Petri and {Schreiber}, Andreas and {Chatterjee}, Piyali and {K{\"a}pyl{\"a}}, Maarit and {Li}, Xiang-Yu and {Kr{\"u}ger}, Jonas and {Aarnes}, J{\o}rgen and {Sarson}, Graeme and {Oishi}, Jeffrey and {Schober}, Jennifer and {Plasson}, Rapha{\"e}l and {Sandin}, Christer and {Karchniwy}, Ewa and {Rodrigues}, Luiz and {Hubbard}, Alexander and {Guerrero}, Gustavo and {Snodin}, Andrew and {Losada}, Illa and {Pekkil{\"a}}, Johannes and {Qian}, Chengeng},
    collaboration = "Pencil Code",
    title = "{The Pencil Code, a modular MPI code for partial differential equations and particles: multipurpose and multiuser-maintained}",
    eprint = "2009.08231",
    archivePrefix = "arXiv",
    primaryClass = "astro-ph.IM",
    reportNumber = "NORDITA-2020-087",
    doi = "10.21105/joss.02807",
    journal = "J. Open Source Softw.",
    volume = "6",
    number = "58",
    pages = "2807",
    year = "2021"
}

@article{Guth:1980zm,
  author = "Guth, Alan H.",
  title = "{The Inflationary Universe: A Possible Solution to the Horizon and Flatness Problems}",
  journal = "Phys. Rev. D",
  volume = "23",
  year = "1981",
  pages = "347--356",
  doi = "10.1103/PhysRevD.23.347"
}

@article{Linde:1981mu,
  author = "Linde, A. D.",
  title = "{A New Inflationary Universe Scenario: A Possible Solution of the Horizon, Flatness, Homogeneity, Isotropy and Primordial Monopole Problems}",
  journal = "Phys. Lett. B",
  volume = "108",
  year = "1982",
  pages = "389--393",
  doi = "10.1016/0370-2693(82)91219-9"
}

@article{Albrecht:1982wi,
  author = "Albrecht, Andreas and Steinhardt, Paul J.",
  title = "{Cosmology for Grand Unified Theories with Radiatively Induced Symmetry Breaking}",
  journal = "Phys. Rev. Lett.",
  volume = "48",
  year = "1982",
  pages = "1220--1223",
  doi = "10.1103/PhysRevLett.48.1220"
}

@article{Freese:1990rb,
  author = "Freese, Katherine and Frieman, Joshua A. and Olinto, Angela V.",
  title = "{Natural inflation with pseudo - Nambu-Goldstone bosons}",
  journal = "Phys. Rev. Lett.",
  volume = "65",
  year = "1990",
  pages = "3233--3236",
  doi = "10.1103/PhysRevLett.65.3233"
}

@article{Svrcek:2006yi,
  author = "Svrcek, Peter and Witten, Edward",
  title = "{Axions In String Theory}",
  journal = "JHEP",
  volume = "06",
  year = "2006",
  pages = "051",
  doi = "10.1088/1126-6708/2006/06/051",
  eprint = "hep-th/0605206"
}

@article{Anber:2009ua,
  author = "Anber, Mohamed M. and Sorbo, Lorenzo",
  title = "{Naturally inflating on steep potentials through electromagnetic dissipation}",
  journal = "Phys. Rev. D",
  volume = "81",
  year = "2010",
  pages = "043534",
  doi = "10.1103/PhysRevD.81.043534",
  eprint = "0908.4089"
}

@article{Barnaby:2011qe,
  author = "Barnaby, Neil and Namba, Ryo and Peloso, Marco",
  title = "{Phenomenology of a pseudo-scalar inflaton: Naturally large non-Gaussianity}",
  journal = "JCAP",
  volume = "04",
  year = "2011",
  pages = "009",
  doi = "10.1088/1475-7516/2011/04/009",
  eprint = "1102.4333"
}

@article{Kluger:1991ib,
  author = "Kluger, Y. and Eisenberg, J. M. and Svetitsky, B. and Cooper, Fred and Mottola, Emil",
  title = "{Fermion pair production in a strong electric field}",
  journal = "Phys. Rev. D",
  volume = "45",
  year = "1992",
  pages = "4659--4671",
  doi = "10.1103/PhysRevD.45.4659"
}

@article{Adshead:2015pva,
    author = "Adshead, Peter and Giblin, John T. and Scully, Timothy R. and Sfakianakis, Evangelos I.",
    title = "{Gauge-preheating and the end of axion inflation}",
    eprint = "1502.06506",
    archivePrefix = "arXiv",
    primaryClass = "astro-ph.CO",
    doi = "10.1088/1475-7516/2015/12/034",
    journal = "JCAP",
    volume = "12",
    pages = "034",
    year = "2015"
}

@article{Adshead:2016iae,
    author = "Adshead, Peter and Giblin, John T. and Scully, Timothy R. and Sfakianakis, Evangelos I.",
    title = "{Magnetogenesis from axion inflation}",
    eprint = "1606.08474",
    archivePrefix = "arXiv",
    primaryClass = "astro-ph.CO",
    doi = "10.1088/1475-7516/2016/10/039",
    journal = "JCAP",
    volume = "10",
    pages = "039",
    year = "2016"
}

@article{MAGIC:2022piy,
    author = "Acciari, V. A. and others",
    collaboration = "MAGIC",
    title = "{A lower bound on intergalactic magnetic fields from time variability of 1ES 0229+200 from MAGIC and Fermi/LAT observations}",
    eprint = "2210.03321",
    archivePrefix = "arXiv",
    primaryClass = "astro-ph.HE",
    doi = "10.1051/0004-6361/202244126",
    journal = "Astron. Astrophys.",
    volume = "670",
    pages = "A145",
    year = "2023"
}

@article{Sorbo:2011rz,
    author = "Sorbo, Lorenzo",
    title = "{Parity violation in the Cosmic Microwave Background from a pseudoscalar inflaton}",
    eprint = "1101.1525",
    archivePrefix = "arXiv",
    primaryClass = "astro-ph.CO",
    doi = "10.1088/1475-7516/2011/06/003",
    journal = "JCAP",
    volume = "06",
    pages = "003",
    year = "2011"
}

@article{Cook:2013xea,
    author = "Cook, Jessica L. and Sorbo, Lorenzo",
    title = "{An inflationary model with small scalar and large tensor nongaussianities}",
    eprint = "1307.7077",
    archivePrefix = "arXiv",
    primaryClass = "astro-ph.CO",
    doi = "10.1088/1475-7516/2013/11/047",
    journal = "JCAP",
    volume = "11",
    pages = "047",
    year = "2013"
}

@article{Bastero-Gil:2022fme,
    author = "Bastero-Gil, Mar and Manso, Ant\'onio Torres",
    title = "{Parity violating gravitational waves at the end of inflation}",
    eprint = "2209.15572",
    archivePrefix = "arXiv",
    primaryClass = "gr-qc",
    doi = "10.1088/1475-7516/2023/08/001",
    journal = "JCAP",
    volume = "08",
    pages = "001",
    year = "2023"
}

@article{Garcia-Bellido:2023ser,
    author = "Garcia-Bellido, Juan and Papageorgiou, Alexandros and Peloso, Marco and Sorbo, Lorenzo",
    title = "{A flashing beacon in axion inflation: recurring bursts of gravitational waves in the strong backreaction regime}",
    eprint = "2303.13425",
    archivePrefix = "arXiv",
    primaryClass = "astro-ph.CO",
    doi = "10.1088/1475-7516/2024/01/034",
    journal = "JCAP",
    volume = "01",
    pages = "034",
    year = "2024"
}

@article{Garretson:1992vt,
    author = "Garretson, W. Daniel and Field, George B. and Carroll, Sean M.",
    title = "{Primordial magnetic fields from pseudoGoldstone bosons}",
    eprint = "hep-ph/9209238",
    archivePrefix = "arXiv",
    reportNumber = "PRINT-92-0448 (CFA,CAMBRIDGE), CFA-3507",
    doi = "10.1103/PhysRevD.46.5346",
    journal = "Phys. Rev. D",
    volume = "46",
    pages = "5346--5351",
    year = "1992"
}

@article{Anber:2006xt,
    author = "Anber, Mohamed M. and Sorbo, Lorenzo",
    title = "{N-flationary magnetic fields}",
    eprint = "astro-ph/0606534",
    archivePrefix = "arXiv",
    doi = "10.1088/1475-7516/2006/10/018",
    journal = "JCAP",
    volume = "10",
    pages = "018",
    year = "2006"
}

@article{Durrer:2023rhc,
    author = "Durrer, R. and Sobol, O. and Vilchinskii, S.",
    title = "{Backreaction from gauge fields produced during inflation}",
    eprint = "2303.04583",
    archivePrefix = "arXiv",
    primaryClass = "gr-qc",
    reportNumber = "MS-TP-23-06",
    doi = "10.1103/PhysRevD.108.043540",
    journal = "Phys. Rev. D",
    volume = "108",
    number = "4",
    pages = "043540",
    year = "2023"
}

@article{Adshead:2023mvt,
    author = "Adshead, Peter and Giblin, John T. and Grutkoski, Ryn and Weiner, Zachary J.",
    title = "{Gauge preheating with full general relativity}",
    eprint = "2311.01504",
    archivePrefix = "arXiv",
    primaryClass = "astro-ph.CO",
    doi = "10.1088/1475-7516/2024/03/017",
    journal = "JCAP",
    volume = "03",
    pages = "017",
    year = "2024"
}

@article{Adshead:2019igv, 
    author = "Adshead, Peter and Giblin, John T. and Pieroni, Mauro and Weiner, Zachary J.",
    title = "{Constraining Axion Inflation with Gravitational Waves across 29 Decades in Frequency}",
    eprint = "1909.12843",
    archivePrefix = "arXiv",
    primaryClass = "astro-ph.CO",
    doi = "10.1103/PhysRevLett.124.171301",
    journal = "Phys. Rev. Lett.",
    volume = "124",
    number = "17",
    pages = "171301",
    year = "2020"
}

@article{Adshead:2019lbr,
    author = "Adshead, Peter and Giblin, John T. and Pieroni, Mauro and Weiner, Zachary J.",
    title = "{Constraining axion inflation with gravitational waves from preheating}",
    eprint = "1909.12842",
    archivePrefix = "arXiv",
    primaryClass = "astro-ph.CO",
    doi = "10.1103/PhysRevD.101.083534",
    journal = "Phys. Rev. D",
    volume = "101",
    number = "8",
    pages = "083534",
    year = "2020"
}

@article{Caravano:2021bfn,
    author = "Caravano, Angelo and Komatsu, Eiichiro and Lozanov, Kaloian D. and Weller, Jochen",
    title = "{Lattice simulations of Abelian gauge fields coupled to axions during inflation}",
    eprint = "2110.10695",
    archivePrefix = "arXiv",
    primaryClass = "astro-ph.CO",
    doi = "10.1103/PhysRevD.105.123530",
    journal = "Phys. Rev. D",
    volume = "105",
    number = "12",
    pages = "123530",
    year = "2022"
}

@article{Caravano:2022epk,
    author = "Caravano, Angelo and Komatsu, Eiichiro and Lozanov, Kaloian D. and Weller, Jochen",
    title = "{Lattice simulations of axion-U(1) inflation}",
    eprint = "2204.12874",
    archivePrefix = "arXiv",
    primaryClass = "astro-ph.CO",
    doi = "10.1103/PhysRevD.108.043504",
    journal = "Phys. Rev. D",
    volume = "108",
    number = "4",
    pages = "043504",
    year = "2023"
}

@article{Caravano:2024xsb,
    author = "Caravano, Angelo and Peloso, Marco",
    title = "{Unveiling the nonlinear dynamics of a rolling axion during inflation}",
    eprint = "2407.13405",
    archivePrefix = "arXiv",
    primaryClass = "astro-ph.CO",
    doi = "10.1088/1475-7516/2025/01/104",
    journal = "JCAP",
    volume = "01",
    pages = "104",
    year = "2025"
}

@article{Figueroa:2024rkr,
    author = "Figueroa, Daniel G. and Lizarraga, Joanes and Loayza, Nicol\'as and Urio, Ander and Urrestilla, Jon",
    title = "{Nonlinear dynamics of axion inflation: A detailed lattice study}",
    eprint = "2411.16368",
    archivePrefix = "arXiv",
    primaryClass = "astro-ph.CO",
    doi = "10.1103/PhysRevD.111.063545",
    journal = "Phys. Rev. D",
    volume = "111",
    number = "6",
    pages = "063545",
    year = "2025"
}

@article{Sauter:1931zz,
    author = "Sauter, Fritz",
    title = "{Uber das Verhalten eines Elektrons im homogenen elektrischen Feld nach der relativistischen Theorie Diracs}",
    doi = "10.1007/BF01339461",
    journal = "Z. Phys.",
    volume = "69",
    pages = "742--764",
    year = "1931"
}

@article{Heisenberg:1936nmg,
    author = "Heisenberg, W. and Euler, H.",
    title = "{Consequences of Dirac's theory of positrons}",
    eprint = "physics/0605038",
    archivePrefix = "arXiv",
    doi = "10.1007/BF01343663",
    journal = "Z. Phys.",
    volume = "98",
    number = "11-12",
    pages = "714--732",
    year = "1936"
}

@article{Fujita:2015iga,
    author = "Fujita, Tomohiro and Namba, Ryo and Tada, Yuichiro and Takeda, Naoyuki and Tashiro, Hiroyuki",
    title = "{Consistent generation of magnetic fields in axion inflation models}",
    eprint = "1503.05802",
    archivePrefix = "arXiv",
    primaryClass = "astro-ph.CO",
    reportNumber = "IPMU-15-0029, ICRR-REPORT-699-2014-25",
    doi = "10.1088/1475-7516/2015/05/054",
    journal = "JCAP",
    volume = "05",
    pages = "054",
    year = "2015"
}

@article{Gorbar:2021rlt,
    author = "Gorbar, E. V. and Schmitz, K. and Sobol, O. O. and Vilchinskii, S. I.",
    title = "{Gauge-field production during axion inflation in the gradient expansion formalism}",
    eprint = "2109.01651",
    archivePrefix = "arXiv",
    primaryClass = "hep-ph",
    reportNumber = "CERN-TH-2021-128",
    doi = "10.1103/PhysRevD.104.123504",
    journal = "Phys. Rev. D",
    volume = "104",
    number = "12",
    pages = "123504",
    year = "2021"
}

@article{Gorbar:2023zla,
    author = "Gorbar, E. V. and Momot, A. I. and Prikhodko, O. O. and Teslyk, O. M.",
    title = "{Hydrodynamical approach to chirality production during axion inflation}",
    eprint = "2311.07429",
    archivePrefix = "arXiv",
    primaryClass = "hep-ph",
    doi = "10.1103/PhysRevD.109.023536",
    journal = "Phys. Rev. D",
    volume = "109",
    number = "2",
    pages = "023536",
    year = "2024"
}

@article{Domcke:2016bkh,
    author = "Domcke, Valerie and Pieroni, Mauro and Bin\'etruy, Pierre",
    title = "{Primordial gravitational waves for universality classes of pseudoscalar inflation}",
    eprint = "1603.01287",
    archivePrefix = "arXiv",
    primaryClass = "astro-ph.CO",
    doi = "10.1088/1475-7516/2016/06/031",
    journal = "JCAP",
    volume = "06",
    pages = "031",
    year = "2016"
}

@article{Garcia-Bellido:2016dkw,
    author = "Garcia-Bellido, Juan and Peloso, Marco and Unal, Caner",
    title = "{Gravitational waves at interferometer scales and primordial black holes in axion inflation}",
    eprint = "1610.03763",
    archivePrefix = "arXiv",
    primaryClass = "astro-ph.CO",
    reportNumber = "IFT-UAM-CSIC-16-100, UMN-TH-3607-16",
    doi = "10.1088/1475-7516/2016/12/031",
    journal = "JCAP",
    volume = "12",
    pages = "031",
    year = "2016"
}

@article{Corba:2024tfz,
    author = "Corb\`a, Sofia P. and Sorbo, Lorenzo",
    title = "{Correlated scalar perturbations and gravitational waves from axion inflation}",
    eprint = "2403.03338",
    archivePrefix = "arXiv",
    primaryClass = "astro-ph.CO",
    doi = "10.1088/1475-7516/2024/10/024",
    journal = "JCAP",
    volume = "10",
    pages = "024",
    year = "2024"
}

@article{Banerjee:2004df,
    author = "Banerjee, Robi and Jedamzik, Karsten",
    title = "{The Evolution of cosmic magnetic fields: From the very early universe, to recombination, to the present}",
    eprint = "astro-ph/0410032",
    archivePrefix = "arXiv",
    doi = "10.1103/PhysRevD.70.123003",
    journal = "Phys. Rev. D",
    volume = "70",
    pages = "123003",
    year = "2004"
}

@misc{PC,
        title = {The Pencil Code.  DOI:10.5281/zenodo.2315093.  \href{https://github.com/pencil-code}{https://github.com/pencil-code}},
        year = "2026"
}

@article{Schwinger:1951nm,
    author = "Schwinger, Julian S.",
    editor = "Milton, K. A.",
    title = "{On gauge invariance and vacuum polarization}",
    doi = "10.1103/PhysRev.82.664",
    journal = "Phys. Rev.",
    volume = "82",
    pages = "664--679",
    year = "1951"
}

@article{Iarygina:2025ncl,
    author = "Iarygina, Oksana and Sfakianakis, Evangelos I. and Brandenburg, Axel",
    title = "{Schwinger effect in axion inflation on a lattice}",
    eprint = "2506.20538",
    archivePrefix = "arXiv",
    primaryClass = "astro-ph.CO",
    reportNumber = "NORDITA-2025-030",
    month = "6",
    year = "2025"
}

@article{ParticleDataGroup:2014cgo,
    author = "Olive, K. A. and others",
    collaboration = "Particle Data Group",
    title = "{Review of Particle Physics}",
    doi = "10.1088/1674-1137/38/9/090001",
    journal = "Chin. Phys. C",
    volume = "38",
    pages = "090001",
    year = "2014"
}

@ARTICLE{Bra+19,
       author = {{Brandenburg}, Axel and {Kahniashvili}, Tina and {Mandal}, Sayan and {Pol}, Alberto Roper and {Tevzadze}, Alexander G. and {Vachaspati}, Tanmay},
        title = "{Dynamo effect in decaying helical turbulence}",
      journal = {Phys. Rev. Fluids},
         year = 2019,
        month = feb,
       volume = {4},
       number = {2},
          eid = {024608},
        pages = {024608},
          doi = {10.1103/PhysRevFluids.4.024608},
archivePrefix = {arXiv},
       eprint = {1710.01628},
 primaryClass = {physics.flu-dyn},
       adsurl = {https://ui.adsabs.harvard.edu/abs/2019PhRvF...4b4608B}
}

@ARTICLE{BN25,
       author = {{Brandenburg}, Axel and {Ntormousi}, Evangelia},
        title = "{Magnetic Field Amplification during a Turbulent Collapse}",
      journal = {\apj},
         year = 2025,
        month = sep,
       volume = {990},
       number = {2},
          eid = {223},
        pages = {223},
          doi = {10.3847/1538-4357/adf725},
archivePrefix = {arXiv},
       eprint = {2505.02885},
 primaryClass = {astro-ph.GA},
       adsurl = {https://ui.adsabs.harvard.edu/abs/2025ApJ...990..223B}
}

@INPROCEEDINGS{Jones08,
       author = {{Jones}, Chris A.},
        title = "{Course 2 Dynamo theory}",
    booktitle = {Les Houches},
         year = 2008,
       series = {Les Houches},
       volume = {88},
        month = jan,
        pages = {45-135},
          doi = {10.1016/S0924-8099(08)80006-6},
       adsurl = {https://ui.adsabs.harvard.edu/abs/2008LHouc..88...45J}
}

@ARTICLE{Reppin+Banerjee17,
       author = {{Reppin}, Johannes and {Banerjee}, Robi},
        title = "{Nonhelical turbulence and the inverse transfer of energy: A parameter study}",
      journal = {\pre},
         year = 2017,
        month = nov,
       volume = {96},
       number = {5},
          eid = {053105},
        pages = {053105},
          doi = {10.1103/PhysRevE.96.053105},
archivePrefix = {arXiv},
       eprint = {1708.07717},
 primaryClass = {astro-ph.CO},
       adsurl = {https://ui.adsabs.harvard.edu/abs/2017PhRvE..96e3105R}
}

@ARTICLE{Trivedi+18,
       author = {{Trivedi}, Pranjal and {Reppin}, Johannes and {Chluba}, Jens and {Banerjee}, Robi},
        title = "{Magnetic heating across the cosmological recombination era: results from 3D MHD simulations}",
      journal = {\mnras},
         year = 2018,
        month = dec,
       volume = {481},
       number = {3},
        pages = {3401-3422},
          doi = {10.1093/mnras/sty1757},
archivePrefix = {arXiv},
       eprint = {1805.05315},
 primaryClass = {astro-ph.CO},
       adsurl = {https://ui.adsabs.harvard.edu/abs/2018MNRAS.481.3401T}
}

@article{Christensson:2000sp,
        archiveprefix = {arXiv},
        author = {Christensson, Mattias and Hindmarsh, Mark and Brandenburg, Axel},
        doi = {10.1103/PhysRevE.64.056405},
        eprint = {astro-ph/0011321},
        journal = {Phys. Rev. E},
        pages = {056405},
        title = {{Inverse cascade in decaying 3-D magnetohydrodynamic turbulence}},
        volume = {64},
        year = {2001}}

@ARTICLE{BHKRS21,
       author = {{Brandenburg}, Axel and {He}, Yutong and {Kahniashvili}, Tina and {Rheinhardt}, Matthias and {Schober}, Jennifer},
        title = "{Relic Gravitational Waves from the Chiral Magnetic Effect}",
      journal = {\apj},
         year = 2021,
        month = apr,
       volume = {911},
       number = {2},
          eid = {110},
        pages = {110},
          doi = {10.3847/1538-4357/abe4d7},
archivePrefix = {arXiv},
       eprint = {2101.08178},
 primaryClass = {astro-ph.CO},
       adsurl = {https://ui.adsabs.harvard.edu/abs/2021ApJ...911..110B}
}

@ARTICLE{BS21,
       author = {{Brandenburg}, Axel and {Sharma}, Ramkishor},
        title = "{Simulating Relic Gravitational Waves from Inflationary Magnetogenesis}",
      journal = {\apj},
         year = 2021,
        month = oct,
       volume = {920},
       number = {1},
          eid = {26},
        pages = {26},
          doi = {10.3847/1538-4357/ac1599},
archivePrefix = {arXiv},
       eprint = {2106.03857},
 primaryClass = {astro-ph.CO},
       adsurl = {https://ui.adsabs.harvard.edu/abs/2021ApJ...920...26B}
}

@ARTICLE{BHS21,
       author = {{Brandenburg}, Axel and {He}, Yutong and {Sharma}, Ramkishor},
        title = "{Simulations of Helical Inflationary Magnetogenesis and Gravitational Waves}",
      journal = {\apj},
         year = 2021,
        month = dec,
       volume = {922},
       number = {2},
          eid = {192},
        pages = {192},
          doi = {10.3847/1538-4357/ac20d9},
archivePrefix = {arXiv},
       eprint = {2107.12333},
 primaryClass = {astro-ph.CO},
       adsurl = {https://ui.adsabs.harvard.edu/abs/2021ApJ...922..192B}
}

@article{Jedamzik:2023rfd,
    author = "Jedamzik, Karsten and Abel, Tom and Ali-Haimoud, Yacine",
    title = "{Cosmic recombination in the presence of primordial magnetic fields}",
    eprint = "2312.11448",
    archivePrefix = "arXiv",
    primaryClass = "astro-ph.CO",
    doi = "10.1088/1475-7516/2025/03/012",
    journal = "JCAP",
    volume = "03",
    pages = "012",
    year = "2025"
}

@article{Jedamzik:2025cax,
    author = "Jedamzik, Karsten and Pogosian, Levon and Abel, Tom",
    title = "{Hints of primordial magnetic fields at recombination and implications for the Hubble tension}",
    eprint = "2503.09599",
    archivePrefix = "arXiv",
    primaryClass = "astro-ph.CO",
    reportNumber = "SCG-2025-01",
    doi = "10.1038/s41550-025-02737-x",
    journal = "Nature Astron.",
    volume = "10",
    number = "2",
    pages = "317--324",
    year = "2026"
}

@article{LISA:2017pwj,
    author = "Amaro-Seoane, Pau and others",
    collaboration = "LISA",
    title = "{Laser Interferometer Space Antenna}",
    eprint = "1702.00786",
    archivePrefix = "arXiv",
    primaryClass = "astro-ph.IM",
    month = "2",
    year = "2017"
}

@article{Caprini:2009yp,
    author = "Caprini, Chiara and Durrer, Ruth and Servant, Geraldine",
    title = "{The stochastic gravitational wave background from turbulence and magnetic fields generated by a first-order phase transition}",
    eprint = "0909.0622",
    archivePrefix = "arXiv",
    primaryClass = "astro-ph.CO",
    doi = "10.1088/1475-7516/2009/12/024",
    journal = "JCAP",
    volume = "12",
    pages = "024",
    year = "2009"
}

@article{RoperPol:2019wvy,
    author = "Roper Pol, Alberto and Mandal, Sayan and Brandenburg, Axel and Kahniashvili, Tina and Kosowsky, Arthur",
    title = "{Numerical simulations of gravitational waves from early-universe turbulence}",
    eprint = "1903.08585",
    archivePrefix = "arXiv",
    primaryClass = "astro-ph.CO",
    reportNumber = "NORDITA-2019-024",
    doi = "10.1103/PhysRevD.102.083512",
    journal = "Phys. Rev. D",
    volume = "102",
    number = "8",
    pages = "083512",
    year = "2020"
}

@ARTICLE{BM99,
       author = {{Biskamp}, Dieter and {M{\"u}ller}, Wolf-Christian},
        title = "{Decay Laws for Three-Dimensional Magnetohydrodynamic Turbulence}",
      journal = {\prl},
         year = 1999,
        month = sep,
       volume = {83},
       number = {11},
        pages = {2195-2198},
          doi = {10.1103/PhysRevLett.83.2195},
archivePrefix = {arXiv},
       eprint = {physics/9903028},
 primaryClass = {physics.flu-dyn},
       adsurl = {https://ui.adsabs.harvard.edu/abs/1999PhRvL..83.2195B}
}

@ARTICLE{Hatori84,
       author = {{Hatori}, Tadatsugu},
        title = "{Kolmogorov-Style Argument for the Decaying Homogeneous MHD Turbulence}",
      journal = {J. Phys. Soc. Jpn},
         year = 1984,
        month = aug,
       volume = {53},
       number = {8},
        pages = {2539},
          doi = {10.1143/JPSJ.53.2539},
       adsurl = {https://ui.adsabs.harvard.edu/abs/1984JPSJ...53.2539H}
}

@ARTICLE{BK17,
   author = {{Brandenburg}, A. and {Kahniashvili}, T.},
    title = "{Classes of Hydrodynamic and Magnetohydrodynamic Turbulent Decay}",
  journal = "\prl",
archivePrefix = "arXiv",
   eprint = {1607.01360},
 primaryClass = "physics.flu-dyn",
     year = 2017,
    month = feb,
   volume = 118,
      eid = {055102},
    pages = {055102},
      doi = {10.1103/PhysRevLett.118.055102},
   adsurl = {http://adsabs.harvard.edu/abs/2017PhRvL.118e5102B}
}

@ARTICLE{Bra+17,
       author = {{Brandenburg}, Axel and {Kahniashvili}, Tina and {Mandal}, Sayan and
        {Pol}, Alberto Roper and {Tevzadze}, Alexander G. and
        {Vachaspati}, Tanmay},
        title = "{Evolution of hydromagnetic turbulence from the electroweak phase
        transition}",
      journal = {\prd},
         year = 2017,
        month = Dec,
       volume = {96},
          eid = {123528},
        pages = {123528},
          doi = {10.1103/PhysRevD.96.123528},
       adsurl = {https://ui.adsabs.harvard.edu/#abs/2017PhRvD..96l3528B}
}

@misc{DATA,
        title = "{Datasets for Primordial turbulence from inflation: a new inflaton-driven turbulent regime; \url{http://norlx65.nordita.org/~brandenb/projects/PrimTurb}}"
}

@ARTICLE{JP20,
       author = {{Jedamzik}, Karsten and {Pogosian}, Levon},
        title = "{Relieving the Hubble Tension with Primordial Magnetic Fields}",
      journal = {\prl},
         year = 2020,
        month = oct,
       volume = {125},
       number = {18},
          eid = {181302},
        pages = {181302},
          doi = {10.1103/PhysRevLett.125.181302},
archivePrefix = {arXiv},
       eprint = {2004.09487},
 primaryClass = {astro-ph.CO},
       adsurl = {https://ui.adsabs.harvard.edu/abs/2020PhRvL.125r1302J}
}

@ARTICLE{Bhat+21,
       author = {{Bhat}, Pallavi and {Zhou}, Muni and {Loureiro}, Nuno F.},
        title = "{Inverse energy transfer in decaying, three-dimensional, non-helical magnetic turbulence due to magnetic reconnection}",
      journal = {\mnras},
         year = 2021,
        month = feb,
       volume = {501},
       number = {2},
        pages = {3074-3087},
          doi = {10.1093/mnras/staa3849},
archivePrefix = {arXiv},
       eprint = {2007.07325},
 primaryClass = {astro-ph.CO},
       adsurl = {https://ui.adsabs.harvard.edu/abs/2021MNRAS.501.3074B}
}

@ARTICLE{HS23,
       author = {{Hosking}, D. N. and {Schekochihin}, A. A.},
        title = "{Cosmic-void observations reconciled with primordial magnetogenesis}",
      journal = {NatCo},
         year = 2023,
        month = nov,
       volume = {14},
          eid = {7523},
        pages = {7523},
          doi = {10.1038/s41467-023-43258-3},
archivePrefix = {arXiv},
       eprint = {2203.03573},
 primaryClass = {astro-ph.CO},
       adsurl = {https://ui.adsabs.harvard.edu/abs/2023NatCo..14.7523H}
}

@ARTICLE{Isochrone26,
       author = {{Brandenburg}, Axel and {Cielo}, Mattia and {Iarygina}, Oksana and {Vazza}, Franco},
        title = "{Isochrones in primordial magnetic field evolution}",
      journal = {arXiv e-prints},
         year = 2026,
        month = jun,
          eid = {arXiv:2606.10863},
        pages = {arXiv:2606.10863},
          doi = {10.48550/arXiv.2606.10863},
archivePrefix = {arXiv},
       eprint = {2606.10863},
 primaryClass = {astro-ph.CO},
       adsurl = {https://ui.adsabs.harvard.edu/abs/2026arXiv260610863B}
}

@ARTICLE{Kahniashvili+08,
       author = {{Kahniashvili}, Tina and {Campanelli}, Leonardo and {Gogoberidze}, Grigol and {Maravin}, Yurii and {Ratra}, Bharat},
        title = "{Gravitational radiation from primordial helical inverse cascade magnetohydrodynamic turbulence}",
      journal = {\prd},
         year = 2008,
        month = dec,
       volume = {78},
       number = {12},
          eid = {123006},
        pages = {123006},
          doi = {10.1103/PhysRevD.78.123006},
archivePrefix = {arXiv},
       eprint = {0809.1899},
 primaryClass = {astro-ph},
       adsurl = {https://ui.adsabs.harvard.edu/abs/2008PhRvD..78l3006K}
}

@ARTICLE{Kahniashvili+05,
       author = {{Kahniashvili}, Tina and {Gogoberidze}, Grigol and {Ratra}, Bharat},
        title = "{Polarized Cosmological Gravitational Waves from Primordial Helical Turbulence}",
      journal = {\prl},
         year = 2005,
        month = oct,
       volume = {95},
       number = {15},
          eid = {151301},
        pages = {151301},
          doi = {10.1103/PhysRevLett.95.151301},
archivePrefix = {arXiv},
       eprint = {astro-ph/0505628},
 primaryClass = {astro-ph},
       adsurl = {https://ui.adsabs.harvard.edu/abs/2005PhRvL..95o1301K}
}

@ARTICLE{Caprini+Durrer01,
       author = {{Caprini}, Chiara and {Durrer}, Ruth},
        title = "{Gravitational wave production: A strong constraint on primordial magnetic fields}",
      journal = {\prd},
         year = 2001,
        month = dec,
       volume = {65},
       number = {2},
          eid = {023517},
        pages = {023517},
          doi = {10.1103/PhysRevD.65.023517},
archivePrefix = {arXiv},
       eprint = {astro-ph/0106244},
 primaryClass = {astro-ph},
       adsurl = {https://ui.adsabs.harvard.edu/abs/2001PhRvD..65b3517C}
}

@ARTICLE{Caprini+Durrer06,
       author = {{Caprini}, Chiara and {Durrer}, Ruth},
        title = "{Gravitational waves from stochastic relativistic sources: Primordial turbulence and magnetic fields}",
      journal = {\prd},
         year = 2006,
        month = sep,
       volume = {74},
       number = {6},
          eid = {063521},
        pages = {063521},
          doi = {10.1103/PhysRevD.74.063521},
archivePrefix = {arXiv},
       eprint = {astro-ph/0603476},
 primaryClass = {astro-ph},
       adsurl = {https://ui.adsabs.harvard.edu/abs/2006PhRvD..74f3521C}
}

@ARTICLE{Bra+21,
       author = {{Brandenburg}, Axel and {Clarke}, Emma and {He}, Yutong and {Kahniashvili}, Tina},
        title = "{Can we observe the QCD phase transition-generated gravitational waves through pulsar timing arrays?}",
      journal = {\prd},
         year = 2021,
        month = aug,
       volume = {104},
       number = {4},
          eid = {043513},
        pages = {043513},
          doi = {10.1103/PhysRevD.104.043513},
archivePrefix = {arXiv},
       eprint = {2102.12428},
 primaryClass = {astro-ph.CO},
       adsurl = {https://ui.adsabs.harvard.edu/abs/2021PhRvD.104d3513B}
}

@ARTICLE{Neronov+21,
       author = {{Neronov}, Andrii and {Roper Pol}, Alberto and {Caprini}, Chiara and {Semikoz}, Dmitri},
        title = "{NANOGrav signal from magnetohydrodynamic turbulence at the QCD phase transition in the early Universe}",
      journal = {\prd},
         year = 2021,
        month = feb,
       volume = {103},
       number = {4},
          eid = {L041302},
        pages = {L041302},
          doi = {10.1103/PhysRevD.103.L041302},
archivePrefix = {arXiv},
       eprint = {2009.14174},
 primaryClass = {astro-ph.CO},
       adsurl = {https://ui.adsabs.harvard.edu/abs/2021PhRvD.103d1302N}
}

@ARTICLE{Schiff+Venumadhav25,
       author = {{Schiff}, Jonathan and {Venumadhav}, Tejaswi},
        title = "{Primordial magnetic fields and modified recombination histories}",
      journal = {arXiv e-prints},
         year = 2025,
        month = jun,
          eid = {arXiv:2506.16517},
        pages = {arXiv:2506.16517},
          doi = {10.48550/arXiv.2506.16517},
archivePrefix = {arXiv},
       eprint = {2506.16517},
 primaryClass = {astro-ph.CO},
       adsurl = {https://ui.adsabs.harvard.edu/abs/2025arXiv250616517S}
}

@ARTICLE{Paoletti+22,
       author = {{Paoletti}, D. and {Chluba}, J. and {Finelli}, F. and {Rubi{\~n}o-Mart{\'\i}n}, J.~A.},
        title = "{Constraints on primordial magnetic fields from their impact on the ionization history with Planck 2018}",
      journal = {\mnras},
         year = 2022,
        month = dec,
       volume = {517},
       number = {3},
        pages = {3916-3927},
          doi = {10.1093/mnras/stac2947},
archivePrefix = {arXiv},
       eprint = {2204.06302},
 primaryClass = {astro-ph.CO},
       adsurl = {https://ui.adsabs.harvard.edu/abs/2022MNRAS.517.3916P}
}

@ARTICLE{Blunier+26,
       author = {{Blunier}, Jeffrey and {Neronov}, Andrii and {Semikoz}, Dmitri},
        title = "{Revision of conservative lower bound on the intergalactic magnetic field from Fermi and Cherenkov telescope observations of extreme blazars}",
      journal = {\aap},
         year = 2026,
        month = jun,
       volume = {710},
          eid = {A254},
        pages = {A254},
          doi = {10.1051/0004-6361/202658891},
archivePrefix = {arXiv},
       eprint = {2506.22285},
 primaryClass = {astro-ph.HE},
       adsurl = {https://ui.adsabs.harvard.edu/abs/2026A&A...710A.254B}
}

@ARTICLE{Vazza+17,
       author = {{Vazza}, F. and {Br{\"u}ggen}, M. and {Gheller}, C. and {Hackstein}, S. and {Wittor}, D. and {Hinz}, P.~M.},
        title = "{Simulations of extragalactic magnetic fields and of their observables}",
      journal = {Classical and Quantum Gravity},
         year = 2017,
        month = dec,
       volume = {34},
       number = {23},
          eid = {234001},
        pages = {234001},
          doi = {10.1088/1361-6382/aa8e60},
archivePrefix = {arXiv},
       eprint = {1711.02669},
 primaryClass = {astro-ph.CO},
       adsurl = {https://ui.adsabs.harvard.edu/abs/2017CQGra..34w4001V}
}

@ARTICLE{Tjemsland+24,
       author = {{Tjemsland}, J. and {Meyer}, M. and {Vazza}, F.},
        title = "{Constraining the Astrophysical Origin of Intergalactic Magnetic Fields}",
      journal = {\apj},
         year = 2024,
        month = mar,
       volume = {963},
       number = {2},
          eid = {135},
        pages = {135},
          doi = {10.3847/1538-4357/ad22dd},
archivePrefix = {arXiv},
       eprint = {2311.04273},
 primaryClass = {astro-ph.HE},
       adsurl = {https://ui.adsabs.harvard.edu/abs/2024ApJ...963..135T}
}

@ARTICLE{Neronov+24,
       author = {{Neronov}, A. and {Vazza}, F. and {Mtchedlidze}, S. and {Carretti}, E.},
        title = "{Revision of upper bound on volume-filling intergalactic magnetic fields with LOFAR}",
      journal = {arXiv e-prints},
         year = 2024,
        month = dec,
          eid = {arXiv:2412.14825},
        pages = {arXiv:2412.14825},
          doi = {10.48550/arXiv.2412.14825},
archivePrefix = {arXiv},
       eprint = {2412.14825},
 primaryClass = {astro-ph.CO},
       adsurl = {https://ui.adsabs.harvard.edu/abs/2024arXiv241214825N}
}

@article{Mtchedlidze:2021bfy,
    author = {Mtchedlidze, Salome and Dom{\'\i}nguez-Fern{\'a}ndez, Paola and Du, Xiaolong and Brandenburg, Axel and Kahniashvili, Tina and O'Sullivan, Shane and Schmidt, Wolfram and Br{\"u}ggen, Marcus},
    title = "{Evolution of Primordial Magnetic Fields during Large-scale Structure Formation}",
    eprint = "2109.13520",
    archivePrefix = "arXiv",
    primaryClass = "astro-ph.CO",
    reportNumber = "NORDITA 2021-082",
    doi = "10.3847/1538-4357/ac5960",
    journal = "Astrophys. J.",
    volume = "929",
    number = "2",
    pages = "127",
    year = "2022"
}

@article{Subramanian:2015lua,
    author = "Subramanian, Kandaswamy",
    title = "{The origin, evolution and signatures of primordial magnetic fields}",
    eprint = "1504.02311",
    archivePrefix = "arXiv",
    primaryClass = "astro-ph.CO",
    doi = "10.1088/0034-4885/79/7/076901",
    journal = "Rept. Prog. Phys.",
    volume = "79",
    number = "7",
    pages = "076901",
    year = "2016"
}

@article{Durrer:2013pga,
    author = "Durrer, Ruth and Neronov, Andrii",
    title = "{Cosmological Magnetic Fields: Their Generation, Evolution and Observation}",
    eprint = "1303.7121",
    archivePrefix = "arXiv",
    primaryClass = "astro-ph.CO",
    doi = "10.1007/s00159-013-0062-7",
    journal = "Astron. Astrophys. Rev.",
    volume = "21",
    pages = "62",
    year = "2013"
}

@article{Vachaspati:2020blt,
    author = "Vachaspati, Tanmay",
    title = "{Progress on cosmological magnetic fields}",
    eprint = "2010.10525",
    archivePrefix = "arXiv",
    primaryClass = "astro-ph.CO",
    doi = "10.1088/1361-6633/ac03a9",
    journal = "Rept. Prog. Phys.",
    volume = "84",
    number = "7",
    pages = "074901",
    year = "2021"
}

@article{Pajer:2013fsa,
    author = "Pajer, Enrico and Peloso, Marco",
    title = "{A review of Axion Inflation in the era of Planck}",
    eprint = "1305.3557",
    archivePrefix = "arXiv",
    primaryClass = "hep-th",
    doi = "10.1088/0264-9381/30/21/214002",
    journal = "Class. Quant. Grav.",
    volume = "30",
    pages = "214002",
    year = "2013"
}

@article{Corba:2025reo,
    author = "Corb{\`a}, Sofia P.",
    title = "{Gravitational wave anisotropies from axion inflation}",
    eprint = "2504.13156",
    archivePrefix = "arXiv",
    primaryClass = "astro-ph.CO",
    doi = "10.1088/1475-7516/2026/01/029",
    journal = "JCAP",
    volume = "01",
    pages = "029",
    year = "2026"
}

@article{vonEckardstein:2025oic,
    author = "von Eckardstein, Richard and Schmitz, Kai and Sobol, Oleksandr",
    title = "{Gravitational waves from axion inflation in the gradient expansion formalism. Part I. Pure axion inflation}",
    eprint = "2508.00798",
    archivePrefix = "arXiv",
    primaryClass = "astro-ph.CO",
    reportNumber = "MS-TP-25-23",
    doi = "10.1007/JHEP01(2026)018",
    journal = "JHEP",
    volume = "01",
    pages = "018",
    year = "2026"
}

@article{vonEckardstein:2025elq,
    author = "von Eckardstein, Richard and Schmitz, Kai and Sobol, Oleksandr",
    title = "{Gravitational waves from axion inflation in the gradient expansion formalism. Part II. Fermionic axion inflation}",
    eprint = "2509.25013",
    archivePrefix = "arXiv",
    primaryClass = "astro-ph.CO",
    reportNumber = "MS-TP-25-27",
    doi = "10.1007/JHEP03(2026)072",
    journal = "JHEP",
    volume = "03",
    pages = "072",
    year = "2026"
}

@article{Felder:2006cc,
    author = "Felder, Gary N. and Kofman, Lev",
    title = "{Nonlinear inflaton fragmentation after preheating}",
    eprint = "hep-ph/0606256",
    archivePrefix = "arXiv",
    doi = "10.1103/PhysRevD.75.043518",
    journal = "Phys. Rev. D",
    volume = "75",
    pages = "043518",
    year = "2007"
}
 
\clearpage
\newpage

\onecolumngrid
\setcounter{secnumdepth}{3}
\setcounter{equation}{0}
\setcounter{figure}{0}
\setcounter{table}{0}
\setcounter{page}{1}
\makeatletter
\renewcommand{\theequation}{S\arabic{equation}}
\renewcommand{\thefigure}{S\arabic{figure}}
\renewcommand{\bibnumfmt}[1]{[#1]}
\renewcommand{\citenumfont}[1]{#1}
\pagestyle{plain}

\begin{center}
\Large{\textbf{Primordial turbulence from inflation: a new inflaton-driven turbulent regime}}\\
\medskip
\textit{Supplemental Material}\\
\medskip
{Oksana Iarygina and Axel Brandenburg}
\end{center}

\subsection{Electromagnetic energy balance and dynamo action}
\begin{figure}[h!]\begin{center}
\includegraphics[width=0.49\columnwidth]{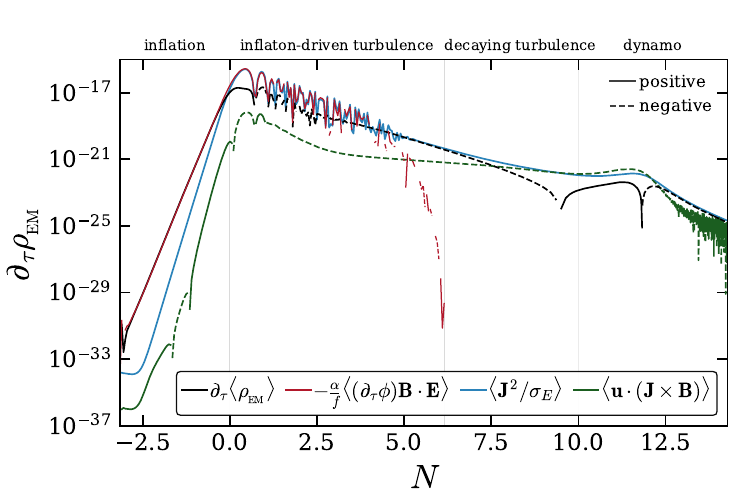}
\includegraphics[width=0.49\columnwidth]{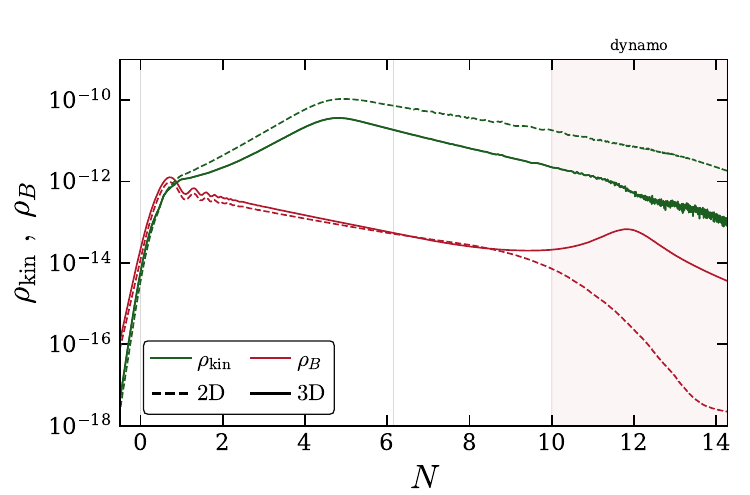}
%Hbdn512alpf90_rho28_ampl1_Gam9o2 
\end{center}\caption[]{\textit{Left:} Evolution of the electromagnetic energy balance $\partial_{\tau}\langle{\rho_\mathrm{EM}}\rangle$ (black) with the inflaton energy-transfer term $-(\alpha/f)\,\langle{(\partial_{\tau} \phi)\, \bm{B}\cdot\bm{E}}\rangle$ (red), electromagnetic energy dissipation $\langle{\bm{J}^2/\sigma_E}\rangle$ (blue), and the work done against the Lorentz force $\langle{\bm{u}\cdot(\bm{J}\times\bm{B})}\rangle$ (green). The balance between the inflaton energy-transfer and dissipation terms during reheating marks the onset of inflaton-driven turbulence. After reheating, the decay of the inflaton  marks the transition to turbulent decay, followed by a short dynamo phase in which the Lorentz-force work exceeds electromagnetic dissipation.
Solid curves correspond to positive values, while dashed curves to negative, with the same color coding.
\textit{Right:} Comparison of the evolution of the magnetic (red curves) and kinetic (green curves) energy densities in the 3D (solid) and 2D (dashed) cases. The increase in magnetic energy occurs only in the 3D case, due to the onset of dynamo action, which cannot be sustained in strictly two-dimensional flows.
}\label{fig:balance}\end{figure}

To identify the mechanism driving the initial inverse cascade and its connection to the inflaton dynamics, we examine the electromagnetic energy balance.
Taking the scalar product of \eqref{eq:Edotconf} with $\bm{E}$, and of the first equation in \eqref{eq:nablaE} with $\bm{B}$, then adding the resulting equations and averaging over the periodic volume, we obtain the electromagnetic energy balance for $\langle\rho_\mathrm{EM}\rangle=\langle\bm{E}^2+\bm{B}^2\rangle/2$.
The resulting equation is
\begin{equation}
\partial_{\tau}\langle{\rho_\mathrm{EM}}\rangle+\frac{\alpha}{f}\langle{(\partial_{\tau} \phi)\, \bm{B}\cdot\bm{E}}\rangle
+\langle{\bm{J}^2/\sigma_E}\rangle+\langle{\bm{u}\cdot(\bm{J}\times\bm{B})}\rangle=0.
\label{eq:balance_drho}
\end{equation}
We examine the evolution of each contribution in the left panel of Figure \ref{fig:balance}. During inflation, $\partial_{\tau}\langle{\rho_\mathrm{EM}}\rangle$ (black) is balanced almost entirely by the inflaton energy-transfer term
$-(\alpha/f)\,\langle{(\partial_{\tau} \phi)\, \bm{B}\cdot\bm{E}}\rangle$ (red). During the subsequent reheating phase, $\partial_{\tau}\langle{\rho_\mathrm{EM}}\rangle$ is predominantly negative,
while the inflaton energy-transfer term is balanced
by electromagnetic energy dissipation via $\langle{\bm{J}^2/\sigma_E}\rangle$ (blue). Both exhibit slowly decaying oscillations and are mostly in phase, with $-(\alpha/f)\,\langle{(\partial_{\tau} \phi)\, \bm{B}\cdot\bm{E}}\rangle$ changing sign. The balance between the inflaton energy-transfer and electromagnetic dissipation coincides with the onset of an inverse cascade, marking a distinct turbulent phase, which we call inflaton-driven turbulence. The phase persists until the end of reheating, when the inflaton $\phi$ decays to zero. Subsequently, the dissipation $\langle{\bm{J}^2/\sigma_E}\rangle$ and the work done against the Lorentz force, $\langle{\bm{u}\cdot(\bm{J}\times\bm{B})}\rangle$ (green), approach each other, marking the onset of turbulent decay. At later times, $\langle{\bm{u}\cdot(\bm{J}\times\bm{B})}\rangle$ exceeds the electromagnetic energy dissipation, leading to a short dynamo phase with magnetic field amplification.

The onset of dynamo action is further demonstrated on the right panel of Figure \ref{fig:balance} by comparing comoving magnetic (red) and kinetic (green) energy densities for 3D (solid lines) and 2D (dashed lines) simulations. We see that the characteristic dynamo bump appears only in the 3D case. This happens because two-dimensional flows cannot sustain a magnetic dynamo \cite{Jones08}.
In 2D, we can express the magnetic field as $\bm{B}=\hat{\bm{z}}B_z+\bm{\nabla}\times(\hat{\bm{z}}A_z)$, where $A_z$ and $B_z$ satisfy
\begin{eqnarray} 
  \partial_\tau A_z+\bm{u}\cdot\bm{\nabla} A_z-\sigma_E^{-1}\nabla^2A_z&=&0,\\
  \partial_\tau B_z+\bm{\nabla}\cdot(\bm{u}B_z)-\sigma_E^{-1}\nabla^2B_z&=&\bm{\nabla}\times(\hat{\bm{z}}A_z)\cdot\bm{\nabla} u_z.
\end{eqnarray}
Both equations are advection--diffusion equations: advection transports $A_z$ and $B_z$ with the flow, while resistive diffusion smooths their spatial gradients,
but that for $A_z$ is source-free, causing the source term in the equation for $B_z$ to vanish.
Although magnetic fields can experience transient amplification through shear or compression \cite{Jones08}, the geometry of a 2D flow does not allow the continuous stretching and folding of magnetic field lines required for self-sustained magnetic field growth. In 3D, there is no analogous reduction to a single scalar $A_z$ obeying a source-free advection--diffusion equation. The additional degrees of freedom of the velocity and magnetic fields enable a sustained transfer of kinetic energy into magnetic energy through the coupling between the fluid and electromagnetic fields, represented by the Lorentz force $\bm{J}\times\bm{B}$ in the momentum equation. Thus, a fully three-dimensional treatment is essential for capturing a genuine turbulent dynamo. The distinct behavior of the 2D and 3D simulations therefore provides a clear signature of the three-dimensional nature of the dynamo instability seen in our work.

\end{document}